\documentclass[trackchanges,twocolumn]{aastex701}

\usepackage{mathrsfs}
\usepackage{xcolor}
\newcommand{\redt}[1]{\textcolor{black}{#1}}

\begin{document}

\title{WASP-43b TESS Phase Curve Mapping: Evidence for a Hot Interior}

\author[orcid=0000-0002-4443-6725]{Maura Lally}
\affiliation{Department of Astronomy and Carl Sagan Institute, Cornell University, 122 Sciences Drive, Ithaca, NY 14853, USA}
\email{ml2289@cornell.edu}  

\author[orcid=0000-0002-8211-6538]{Ryan C. Challener} 
\affiliation{Department of Astronomy and Carl Sagan Institute, Cornell University, 122 Sciences Drive, Ithaca, NY 14853, USA}
\email{rcc276@cornell.edu}

\author[orcid=0000-0002-3052-7116]{Elspeth Lee}
\affiliation{Center for Space and Habitability, University of Bern, Gesellschaftsstrasse 6, CH-3012 Bern, Switzerland}
\email{elspeth.lee@unibe.ch}

\author[orcid=0000-0002-8507-1304]{Nikole Lewis}
\affiliation{Department of Astronomy and Carl Sagan Institute, Cornell University, 122 Sciences Drive, Ithaca, NY 14853, USA}
\email{nikole.lewis@cornell.edu}


\begin{abstract}
 
Phase-curve mapping probes various atmospheric properties depending on wavelength: thermal phase curves are dominated by atmospheric emission, while visible and near-infrared phase curves include reflected light. 
The measured planetary phase curve offset—the longitudinal offset of the hemisphere of peak brightness from the substellar point—is therefore sensitive to differing atmospheric processes in emission- vs. reflection-dominated bands. 
In particular, planets with partial western-dayside reflective cloud coverage show a westward reflected-light phase-curve offset relative to the thermal wavelengths. 
In this work, we analyze \redt{five} sectors of TESS phase curve observations of the hot Jupiter WASP-43b and compare to previous published multi-band analyses to investigate dayside cloud coverage. 
We measure a mid-eclipse depth of \redt{$130 \pm 34$} ppm, indicating excess TESS planetary brightness above expectations based on observed and modeled thermal emission.
Attributing this excess flux to reflection, we estimate a low geometric albedo of \redt{$A_g \approx0.05-0.10$}. 
These findings support previous conclusions that WASP-43b lacks any significant cloud coverage on its dayside. 
Phase-curve mapping reveals a large eastward phase curve offset of \redt{$44 \pm 18$} degrees, significantly eastward of offsets measured at longer wavelengths with JWST, providing evidence for a cloud-free dayside. 
Both the detection of excess emission in the TESS band and the large measured eastward phase-curve offset support the conclusion that WASP-43b has a hot deep atmosphere that is probed at short wavelengths.

\end{abstract}



\section{Introduction} \label{sec:introduction}

Orbital phase curves--measurements of the combined stellar and planetary flux over the duration of an exoplanet's orbit--provide a powerful means of probing atmospheric circulation and energy transport. Phase curves encode longitudinal information about temperature, albedo, and composition on a global scale \citep[e.g.,][]{cowanfujii2024}. Hot Jupiters provide particularly favorable conditions for phase curve studies. Their short orbital periods lead to strong tidal forces that tidally lock the planets' rotation, resulting in permanent day and night sides. Intense stellar irradiation on the dayside produces extreme temperature gradients, leaving the nightside cool enough for cloud formation \citep[e.g.,][]{roman2021, parmentier2021, bell2024, challener2024}. 

General circulation models (GCMs) predict that large temperature gradients between the hot dayside and cool nightside drive eastward equatorial jets, which can advect the hottest region of the atmosphere east of the substellar point \citep[e.g.,][]{showman2009}. Phase curve observations in the infrared (IR) bands are dominated by flux contributions from planetary emission, offering a laboratory to test these models. In these thermal wavelengths, an eastward-shifted hotspot (an eastward phase curve offset) implies efficient advection relative to radiative cooling, whereas a substellar hotspot suggests rapid radiative cooling or strong atmospheric drag \citep{beltz2022, coulombe2023, kennedy2025}. However, IR bands are not sensitive to certain atmospheric characteristics affecting the overall energy budget: the presence of clouds can change the dayside albedo \citep[e.g.,][]{sudarsky2000, burrows2008}, the pressure where irradiation energy is deposited \citep{hengdemory2013}, and cause complex climate feedback depending on their horizontal location \citep{kennedy2025}. In shorter wavelength bands, which are sensitive to reflected light \citep{burrows2008}, cloud coverage can shift the observed longitude of maximum brightness depending on the spatial distribution of the clouds. For example, reflective clouds advected from the nightside onto the morning terminator can produce a westward phase curve offset in optical bands \citep{parmentier2021}. 

The longitudinal information encoded in phase curves can be extracted through brightness mapping \citep[e.g.,][]{knutson2007, hammond2024, challener2024}. Fitting low order spherical harmonic map models to phase curve data provides estimates of the dayside and nightside brightness, the day-night flux contrast, and the phase curve offset. Observing the flux offset in both emission-sensitive long bands and reflection-sensitive short bands is key to connecting observational data to GCMs: comparisons between long and short band brightness temperature maps can be used to disentangle the contributions from thermal hotspot shifts and spatial variations in reflective components \citep{coulombe2025}.  

The hot Jupiter WASP-43b \citep{hellier2011} provides an excellent laboratory for comparative phase curve analysis in both reflection-dominated and emission-dominated wavelengths. The planet has been observed and its heat distribution has been mapped in long emission-sensitive bands using JWST’s MIRI LRS instrument \citep{challener2024, hammond2024} and Spitzer's IRAC \citep{stevenson2017, morello2019, murphy2023}. The planet has also been studied in short reflection-sensitive bands, including phase curve studies using TESS \citep{scandariato2022, arora2024} and HST \citep{stevenson2014}. 

We present a new analysis of currently available TESS phase curve data of WASP-43b, adding to the growing literature on the planet's atmosphere and allowing robust comparison between long and short band maps. The TESS survey observing strategy offers practical advantages as well—continuous 27 day monitoring across multiple sectors offers a long baseline of phase curve data. WASP-43b was observed in \redt{5} TESS sectors, with \redt{4} of them including high-cadence (20s) data. These observations are invaluable for studying the reflective and thermal properties of hot Jupiter atmospheres generally, WASP-43b specifically, and for constraining the dynamics that generate their characteristic hotspot offsets and cloud distributions \citep[e.g.][]{coulombe2025, parmentier2016, parmentier2021}. Section \ref{sec:methods} describes our data acquisition, reduction, and analysis methods in detail. Our results are presented and discussed in Section \ref{sec:results}, with comparisons to previous work in both long and short bands, as well as GCMs in both emission and reflection. We conclude in Section \ref{sec:conclusion}.

\section{Methods} \label{sec:methods}

We analyze \redt{5} sectors of TESS observations of our target WASP-43b, starting with a custom data processing stem to remove long baseline noise which is described in Section \ref{subsec:data processing} and Figure \ref{fig:pre-processing}. We then fit each sector of processed data with \texttt{ThERESA}, detailed in Section \ref{subsec:data analysis} and using fixed planetary parameters shown in Table \ref{table:params_table}. 

\subsection{Data Acquisition}
\label{subsec:data acquisition}

Our target WASP-43 is located within a section of sky which TESS has observed \redt{5} separate times during its flight: sector 9 in year 1, sector 35 in year 3, sector 62 in year 5, sector 89 in year 7, \redt{and sector 100 in year 8}. High cadence (20s) data is available for sectors 35, 62, 89, \redt{and 100}, while sector 9 has only 2-minute cadence data. 

We downloaded the TESS light curves at the highest available cadence for all \redt{five} available sectors of WASP-43b from the Mikulksi Archive for
Space Telescopes (MAST) using the publicly available Light-kurve software package \citep{lightkurve-package}. \redt{All TESS data used in this paper can be accessed via \dataset[doi:10.17909/t36y-d042]{https://doi.org/10.17909/t36y-d042}}. Each sector includes an approximately 27 day lightcurve of the target, with the sectors separated by about two years.

Flux data from MAST is available in two separate levels of processing: Simple Aperture Photometry (SAP) flux, or Pre-search Data Conditioning SAP (PDCSAP) flux. SAP flux is the product of pixel summation of calibrated flux within an optimal aperture, without any further systematics correction. PDCSAP flux is a more processed product, with long-term trends and flux from nearby targets removed from the lightcurve, and points with quality flags discarded (with flags indicating possible anomalies in the data such as Earth/Moon straylight, cosmic ray hits, mechanical issues, and others \citep{kepler_pipeline2010, smith2012, stumpe2012, stumpe2014}). For our analysis we chose to move forward with the PDCSAP flux products in light of the conclusions in \citet{scandariato2022} that the more robust processing is not a source of artificial correlated noise in the processed data.  

\subsection{Data Processing} 
\label{subsec:data processing}

\begin{figure*}[ht!]
\centering
   \includegraphics[width=3.25in]{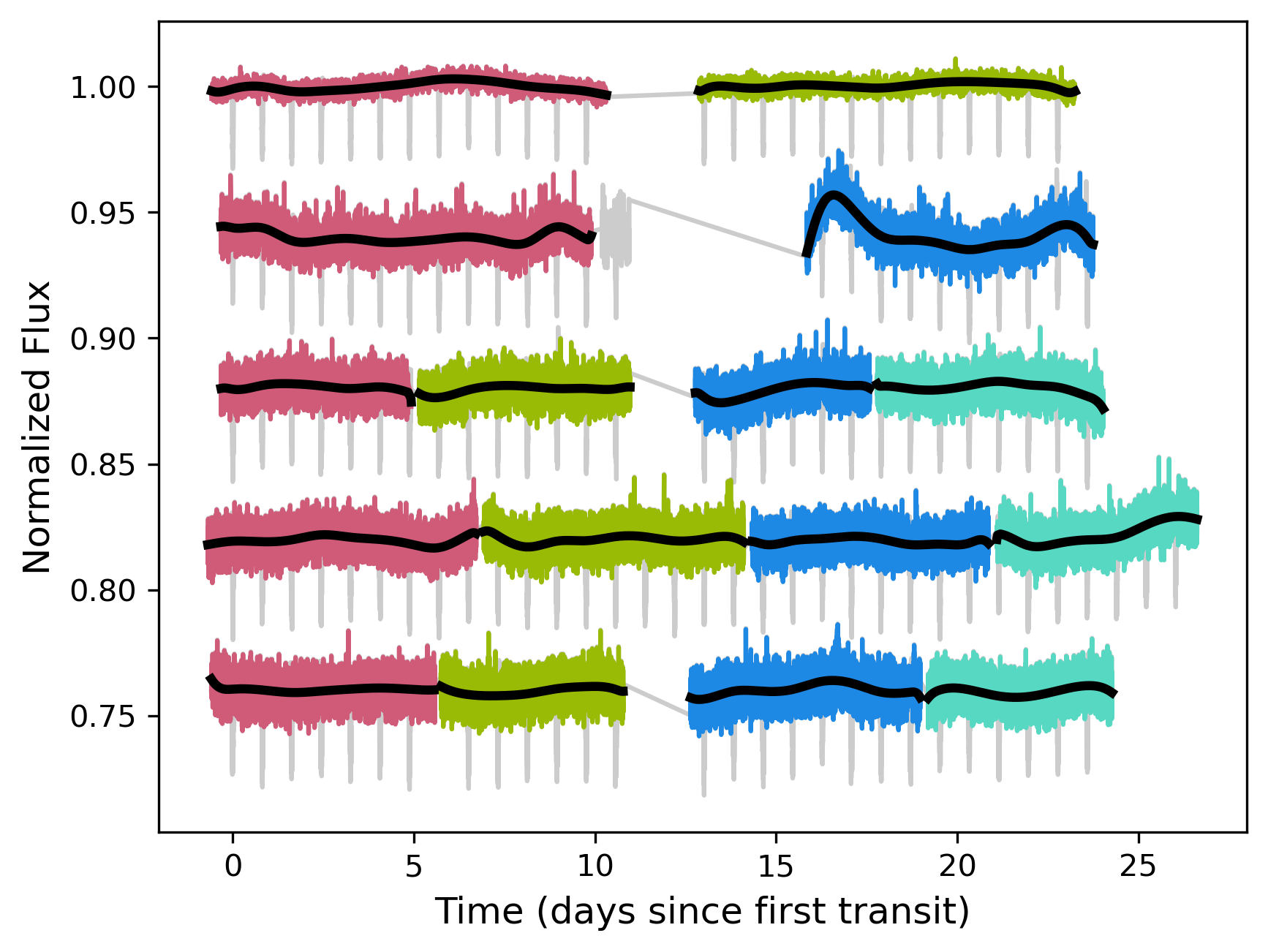}
   \includegraphics[width=3.25in]{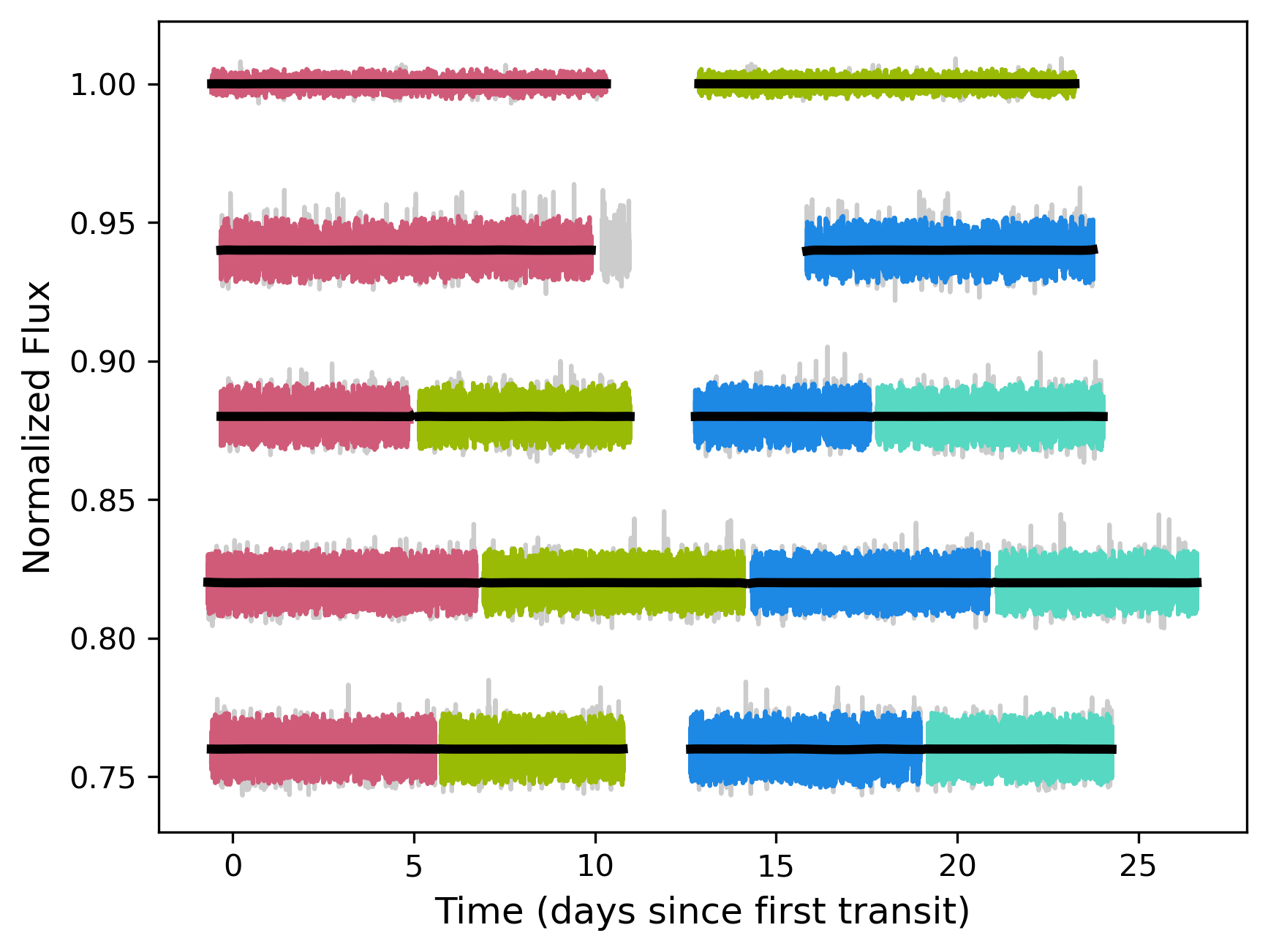}
   \caption{Data processing steps illustrated for each TESS sector, with an offset applied to sectors 35, 62, 89 \redt{and 100} for visibility. In this schematic, all data flagged by the TESS PDCSAP pipeline have been removed. \textbf{Left:} PDCSAP TESS data is shown in gray. In-transit data is removed, and data is split into sub-sections to avoid gaps longer than 3 transit durations. Sub-sections shorter than one orbital period are discarded (this happens in one case, shown in sector 35). Each subset of remaining data (shown in color) is fit with a cubic spline with a knot spacing equal to one orbital period (black).     \textbf{Right:} TESS data (as processed in the left panel) with the spline fit removed is shown in gray. $3\sigma$ outliers of this spline-flattened data are removed (results shown in color). Spline fitting and sigma-clipping steps are iterated until no outliers are identified.}
   \label{fig:pre-processing}
\end{figure*}

We apply custom data reduction steps to the time series data products downloaded from MAST. These custom steps are intended to remove long baseline variations from systematic or stellar sources, while preserving the phase curve and eclipse signal. A schematic overview is shown in Figure \ref{fig:pre-processing}. 

We first ensure all data with quality flags indicating possible anomalies (including cosmic ray hits, stray light detections, detector discontinuities, etc. \citep{smith2012, stumpe2012, stumpe2014}) are removed from the dataset. \redt{These quality flags remove 10\% of the raw TESS data points in our overall data set.} We then remove in-transit data from all sectors of the TESS lightcurve. In-transit data is defined as 1.5x the transit duration centered on the transit time (see Table \ref{table:params_table}); this removes a bit of extra data surrounding the transit, but ensures all in-transit data is removed even in the case of some transit timing variations or long-term orbital decay \citep{sun2018, davoudi2021, bernabo2025} over the span of the approximately 8 years of TESS visits. \redt{Removing the transits excludes 9\% of our overall dataset.}

In order to remove long baseline variations unrelated to our planetary signal, we fit sections of the light curve with a cubic basis spline. To ensure that systematic jumps between TESS orbits would not complicate the spline fitting, we split each sector of data into smaller subsections, enforcing that no single subsection had a gap longer than 3 transit durations. To avoid overfitting and flattening of real phase curve and eclipse signal, we also ensured that no subsection was shorter than a single orbital period. This left us with \redt{16} total subsections of data spanning the \redt{5} original TESS sectors. To process each of these subsections, we follow the method described in \citet{lally2022} and \citet{vanderburg2014}: a cubic spline with a knot spacing equal to one orbital period is fit and removed from the lightcurve, $3\sigma$ outliers are removed, and the spline is re-fit. These steps are iterated until the sigma-clipping step no longer identifies outliers. \redt{In this case, 4\% of our dataset was flagged and removed.}

\redt{In order to ensure the spline detrending does not attenuate true phase curve variations, we repeated the data processing and analysis steps described in Sections \ref{subsec:data processing} and \ref{subsec:data analysis} with a spline knot spacing of two orbital periods, and again with a spacing of 3 orbital periods. The analysis with one-orbit and two-orbit spacing yielded the same phase curve offset, and eclipse depths which differed by $<0.2\sigma$. Spline fitting with 3-orbit knot spacing was unable to properly detrend long-baseline systematics, leaving visible artifacts in the TESS lightcurve data. Following \citet{lally2022}, we use a spacing of one knot per orbit for remaining analysis and results, as it offers the most flexible long-baseline detrending without attenuating the phase curve signal.}

After these processing steps, the lightcurve subsections are reassembled into the original \redt{5} TESS sectors for further analysis. 

\subsection{Data Analysis}
\label{subsec:data analysis}

The \redt{five} sectors of processed TESS data are fit with eclipse mapping code \texttt{ThERESA}\footnote{\url{https://github.com/rychallener/theresa}} \citep{challener2022}, the Three-dimensional Exoplanet Retrieval from Eclipse Spectroscopy of Atmospheres. This package uses principal component analysis to generate “eigencurves” \citep{rauscher2018}, which are a set of orthogonal light curves made from spherical harmonic maps \citep{starry}. The package then uses a least-squares routine to find a linear combination of eigencurves to best fit the observed light curve, followed by a Markov-chain Monte Carlo (MCMC) using \texttt{MC3} \citep{cubillos2017} to estimate uncertainties. \texttt{ThERESA} has been tested through application to synthetic data \citep{challener2022, schlawin2023, challener_rauscher2023, valentine2025} and applied to numerous data sets \citep{coulombe2023, hammond2024, challener2024, valentine2024, schlawin2024b, lally2025}.

\begin{deluxetable}{ccc}[htbp]
\label{table:params_table}
\tablecaption{Fixed Input Parameters for \textsc{ThERESA} Model}
\tablehead{
  \colhead{Parameter} & 
  \colhead{Units}     &
  \colhead{Value}
}
\startdata
Stellar mass $M_s$ & $(M_\odot)$ & $0.6916$ \\
Stellar radius $R_s$ & $(R_\odot)$ & $0.665$ \\
Planet mass $M_p$ & $(M_\odot)$ & $0.001959$ \\
Planet radius $R_p$ & $(R_\odot)$ & $0.1056$ \\
Planet orb. period $p_o$ & (days) & $0.8134741$ \\
Planet rot. period $p_r$ & (days) & $0.8134741$ \\
Eccentricity $e$ &  & $0$ \\
Inclination $i$ & (deg) & $82.155$ \\
Semi-major axis $a$ & (AU) & $0.01504$ \\
Transit time $t_c$$^*$ & (BJD$\_$TDB) & \redt{$2458545.2299$} \\
\enddata
\footnotesize{Values consistent with \citet{challener2024}, with the exception of $t_c$. \\ $^*$ Transit time $t_c$ results from a least-squares fit of all phase-folded TESS lightcurve data to a \texttt{batman} eclipse model with $t_c$ as the only free parameter, other values fixed as stated \citep{batman2015}.} 
\end{deluxetable}

The ThERESA astrophysical model’s functional form is

\begin{equation} \label{eq:sys}
F_{sys}(t) = (1 - B)\left(1+c_0 Y_0^0 + \sum_{i=1}^N c_i E_i\right)
\end{equation}

where $F_{sys}$ is the system flux, $B$ is a normalization constant (fit independently to each TESS data sector), $N$ is the number of eigencurves used, $c_i$ are the eigencurve weights, $E_i$ are the eigencurves, $Y_0^0$ is the uniform-map spherical harmonic component, and the system flux overall is star normalized. The parameter $N$ has a user-defined maximum; eigencurves within the user-defined $l_{max}$ limit are ranked by their variance, and only the $N$ highest-variance eigencurves are used in the final map, a process which is optimized by minimizing the Bayesian information criterion (BIC) \citep{schwarz1978, liddle2007} in accordance with previous mapping work \citep[e.g.][]{challener_rauscher2023, schlawin2023, challener2024, schlawin2024}. 

For this work, we also considered the results of minimizing the Akaike Information Criterion (AIC \citealp{liddle2007}), also discussed in \citet{hammond2024} and \citet{valentine2024}. Both AIC and BIC are model criteria which balance the model’s fit to the data against the model’s complexity; their functional forms are 
\begin{equation}\label{eq:sys}
\begin{split}
AIC & = 2k - 2\ln(L) \\
BIC & = k\ln(n) - 2\ln(L)
\end{split}
\end{equation}
where $k$ is the number of parameters in the model, $L$ is the maximum likelihood, and $n$ is the number of data points in the sample. For large datasets such as the one used in this research, BIC more strictly penalizes model complexity, as the penalty for extra model parameters increases with the size of the dataset. Therefore, while both statistics penalize model complexity, the BIC will prefer simpler models relative to the AIC.

\texttt{ThERESA} creates a measured brightness map and then converts it to a temperature map. The conversion assumes a PHOENIX model stellar spectrum \citep{husser2013} assuming stellar parameters from \citet{esposito2017}, which is integrated over the throughput wavelengths (in this case, the TESS bandpass of 600-1000nm). The planet is also assumed to be a blackbody which, for a range of possible planetary temperatures, is integrated over the bandpass to create a sampling of $T_p(F_p)$. This sampling is then interpolated to the measured planet-to-star flux ratio to create a temperature map associated with the brightness map. Note that this resulting temperature map will have values assuming all planetary flux to be emission, and does not consider the contributions of reflection in the TESS bandpass. Therefore, the brightness temperature values do not correspond to actual planetary temperature, but instead provide an upper limit.

Each TESS data sector is treated as a separate observation of the same target with the same instrument: the mapping model is consistent across all sectors, but a normalization term is allowed to vary between sectors, accounting for stellar or instrument changes in the $\sim2$ year stretches between sector observations. Although our dataset includes significantly more phase curve data than in-eclipse data, and we do not expect a strong mapping signal during eclipse ingress and egress, \texttt{ThERESA} fitting is a worthwhile improvement over simple sinusoidal phase curve fitting. \texttt{ThERESA} ensures a more physically plausible map consistent with the stellar and planetary parameters, enforcing non-negative flux at all longitudes. In addition, the sinusoidal model approach employed in previous analyses \citep[e.g.][]{stevenson2017,scandariato2022} do not capture the eclipse shape, while ThERESA mapping ensures that both the phase curve data and in-eclipse data is consistent with the best-fit map.

\section{Results and Discussion} \label{sec:results}

\begin{figure}[ht!]
\centering
   \includegraphics[width=3.25in]{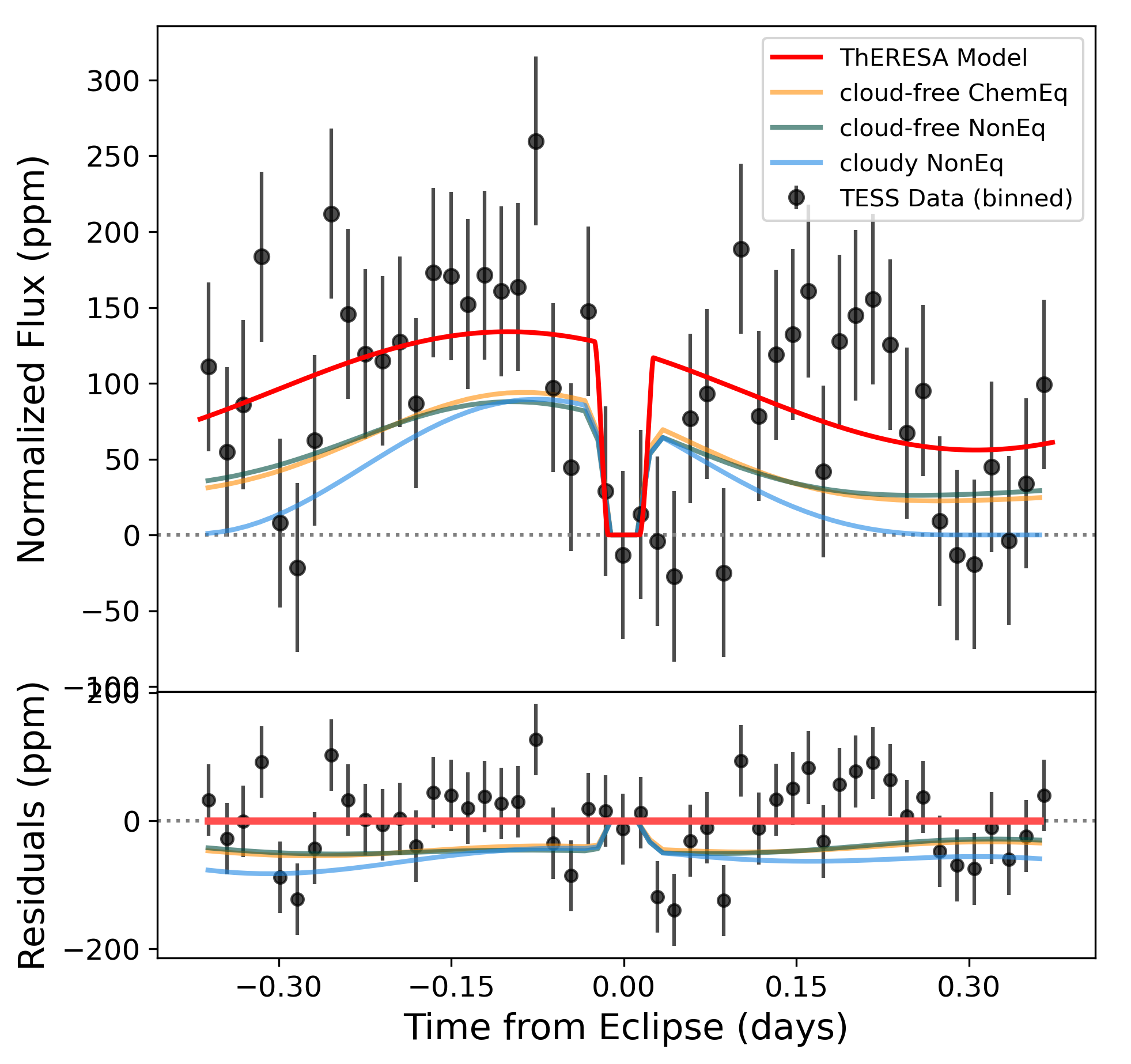}
   \caption{Phase curve \texttt{ThERESA} fit result. Black points show the TESS data, phase folded and binned to 50 data points. Red line shows the $N = 2$ model fit \redt{(reduced $\chi ^2 = 1.008$)}.TESS wavelength phase curves are also shown for the three GCM scenarios described in Section \ref{sec:results}. This direct comparison of our observed phase curve to a sample of emission-only GCMs shows the clear presence of excess flux in the TESS band relative to expectations from modeled emission.}
   \label{fig:lc}
\end{figure}

The result of our phase curve fit is shown in Figure \ref{fig:lc}, which also includes a set of \texttt{Exo-FMS} \citep{lee2021} GCM phase curves modeling the target for comparison. As in the \citet{challener2024} NIRSpec analysis of WASP-43b, our GCMs use fixed planetary parameters from the WASP-43b MIRI LRS analysis published in \citet{bell2024}, and assume an internal temperature of $150$K and solar metallicity. We explore three atmospheric scenarios: (1) ``$cfce$", a cloudless atmosphere assuming chemical equilibrium; (2) ``$cfne$", a cloudless atmosphere with non-equilibrium kinetic chemistry, and (3) ``$cldne$", an atmosphere with tracer clouds assuming non-equilibrium chemistry. For this analysis, we use the \texttt{gCMCRT} code \citep{lee2021_2} to produce white light phase curves in the TESS bandpass for each tested scenario, which are shown in Figure \ref{fig:lc}.

\begin{figure*}[ht!]
\centering
   \includegraphics[width=5.5in]{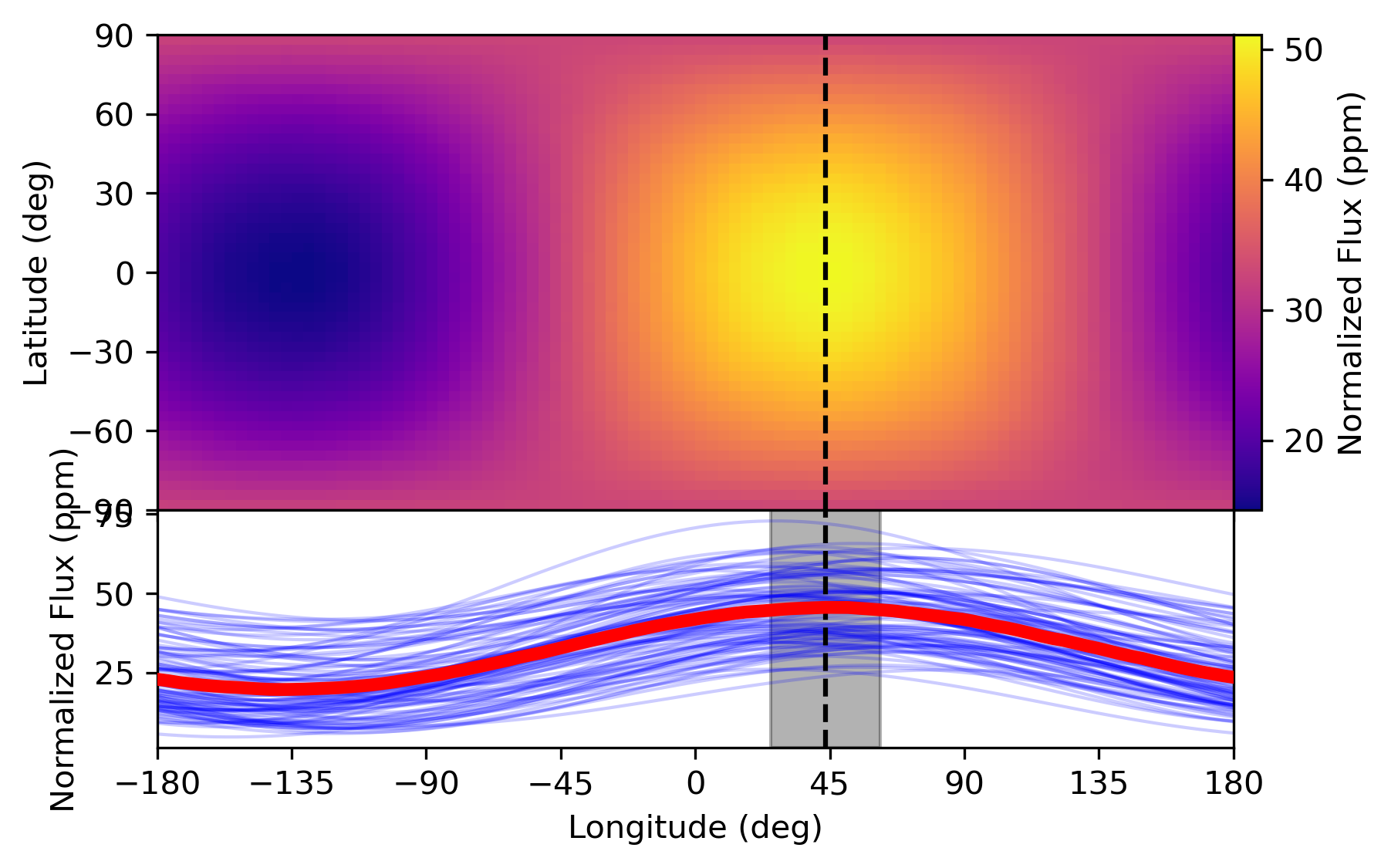}
   \caption{\textbf{Top:} Map of our $N=2$ fit. Vertical black dotted line shows the longitude of peak brightness. Note that while this map shows latitude as well, our $N=2$ fit is not sensitive to latitudinal variations, only longitudinal offsets. \redt{Therefore, the latitudinal stucture shown in the figure is drawn from the spherical harmonic model fitting, and is not driven by the data.} \textbf{Bottom:} Latitudinally averaged flux of the planet. The blue curves correspond to 100 draws from our MCMC posteriors of our $N=2$ map. The red curve corresponds to the median model solution. Shown in black is the longitude of peak flux, with 1$\sigma$ uncertainty shown as the gray shaded region. We present a $3\sigma$ detection of an eastward peak flux offset of \redt{$44\pm18$} degrees.}
   \label{fig:map}
\end{figure*}

We present our \texttt{ThERESA} fit results, as well as derived peak flux offset and geometric albedo, in Table \ref{table:results_table}. Our mapping result is shown in Figure \ref{fig:map}. We discuss model selection in Section \ref{subsec:model selection}, and describe details of deriving geometric albedos using multiple different methods in Section \ref{subsec:eclipse depth and albedo}. We present our peak flux offset result and discuss it in comparison with previously published analyses in Section \ref{subsec: phase curve mapping}. 

\begin{deluxetable}{ccc}[htbp]
\label{table:results_table}
\tablecaption{$l=1, ~N=2$ Model Results}
\tablehead{
  \colhead{Parameter} &
  \colhead{Value}
}
\startdata
\hline
\multicolumn{2}{l}{\texttt{ThERESA} Fit Parameters} \\
\hline
\multicolumn{2}{l}{$~     ~$Mapping Components (ppm):} \\
$c_1$ & \redt{$15.8^{+6.3}_{-6.0}$} \\
$c_2$ & \redt{$16.1^{+7.6}_{-7.6}$} \\
$c_0$ & \redt{$104^{+33}_{-28}$} \\
\multicolumn{2}{l}{$~     ~$TESS Sector Normalization Constants:} \\
$B_1$  & \redt{$1.02^{+0.31}_{-0.35} \times 10^{-4}$} \\
$B_2$  & \redt{$1.10^{+0.32}_{-0.34} \times 10^{-4}$} \\
$B_3$  & \redt{$1.11^{+0.32}_{-0.34} \times 10^{-4}$} \\
$B_4$  & \redt{$1.09^{+0.31}_{-0.34} \times 10^{-4}$} \\
$B_5$  & \redt{$1.08^{+0.33}_{-0.35} \times 10^{-4}$} \\
\hline
\multicolumn{2}{l}{Derived Parameters} \\
\hline
Peak offset ($\phi$, deg) & \redt{$43.6 \pm 18.4$}\\
$A_{g,cfce}$ & \redt{$0.047\pm0.03^{a,b}$} \\
$A_{g,cfne}$ & \redt{$0.053\pm0.03^{a,b}$} \\
$A_{g,cldne}$ & \redt{$0.050\pm0.03^{a,b}$} \\
$A_{g,NIRSpec}$ & \redt{$0.098\pm0.03^b$} \\
\enddata 
\footnotesize{ $^a$Reported uncertainties are based only on the uncertainty in the observed TESS eclipse depth, and assuming zero uncertainty on the GCM eclipse depths. \\

$^b$ \redt{Due to uncertainties introduced in our attempts to remove the thermal phase curve component, these albedo values should be taken as evidence for a low-albedo atmosphere, and not direct albedo measurements.}}
\end{deluxetable}

\subsection{Model Selection}
\label{subsec:model selection}

As discussed in Section \ref{sec:methods}, this analysis considers both BIC and AIC minimization in our model selection. We evaluate the relative likelihood $\mathscr{L}$ of each model according to the expression $\mathscr{L} = e^{-\Delta/2}$, where $\Delta$ is the difference in fit metric (either AIC or BIC) between a given model and the model for which the metric is minimized (the ``best fit"). The BIC minimization prefers a uniform dayside model as a best fit, \redt{finding it significantly preferable to a 2-component model sensitive to longitudinal asymmetry ($\Delta BIC = 13.1$)}. 
However, the AIC minimization strongly prefers a 2-component model, finding it to be over \redt{$40$} times as likely as the uniform model \redt{($\Delta AIC = 7.4$)}. \redt{The 2-component model also unsurprisingly achieves a better $\chi ^2$ than the uniform model ($\Delta \chi ^2 = 11.4$), since it includes the effect of longitudinal structure on the phase curve.}

\redt{Although this work does not conclusively rule out a uniform dayside model, we adopt the 2-component model as our preferred interpretation moving forward. Motivated by the strength of the AIC and $\chi ^2$ preference for the model, as well as the physical plausibility of a phase curve offset, we find it appropriate to move forward with interpretation of this more complex model. The 2-component model is sensitive to longitudinal variations, which is necessary to facilitate comparison to previous phase curve studies.} 

Fit parameters from this model, as well as other derived parameters discussed in this paper, can be found in Table \ref{table:results_table}. 

\subsection{Eclipse Depth and Albedo}
\label{subsec:eclipse depth and albedo}

We find a mid-eclipse planet-to-star flux ratio of \redt{$130 \pm 34$ ppm with a $3.8\sigma$ significance}. Our measured eclipse depth is consistent with the \citet{scandariato2022} analysis of the first two sectors of TESS data ($110\pm50$) as well as the reanalysis of those sectors in \citet{arora2024} ($182^{+64}_{-56}$). 
Figure \ref{fig:lc} compares our phase-curve model with GCMs modeling several atmospheric scenarios for WASP-43b, considering only emission at TESS wavelengths \citep{challener2024, lee2021, lee2021_2}. Our result shows excess dayside flux when compared to both shorter-wave observations and emission-only models. 

In order to extract a geometric albedo from our result, we must first remove our model phase curve's thermal emission component, to find a planet-to-star flux ratio attributable only to reflection \citep{scandariato2022, seager2010}. We report two different approaches to modeling this emission component: first using emission-only GCMs \citep{challener2024, lee2021, lee2021_2}, and second, using IR data to model the contribution from emission at TESS wavelengths. \redt{Both methods are imprecise, and the process of assuming and removing a thermal emission component is prone to introducing uncertainty which is difficult to quantify. The albedo results from both methods should be taken as evidence of a low albedo atmosphere, as opposed to a direct measurement.}

\subsubsection{GCM emission}
\label{subsubsec:GCM emission}

We assume the GCM phase curves discussed in this work represent the entire emission contribution to our observed phase curve, and we compute the geometric albedo by removing this emission contribution from our measured eclipse depth. Our resulting geometric albedos, computed for each of the model scenarios described, are shown in Table \ref{table:results_table}. This approach to modeling the emission component relies on a self-consistent 3D model of the planet which appropriately adjusts the photosphere for wavelength-dependence. That said, this method does not appropriately capture the uncertainty on the true planetary emission in the TESS band, resulting in a likely underestimated uncertainty on our geometric albedo.

Removing the emission contribution from the equilibrium-chemistry cloud-free GCM ($cfce$) results in a reflection-only eclipse depth of \redt{$49.7~$ppm}. Removing emission from the non-equilibrium cloud-free ($cfne$) and cloudy ($cldne$) models respectively results in reflection-only eclipse depths of \redt{$55.9~$ppm and $53.3~$ppm}. These results differ from each other by a maximum of \redt{$18\%$} of our measured eclipse depth uncertainty. Conclusions from \citet{challener2024} suggest that the non-equilibrium cloudy model is, of the three simulated atmospheres considered, the closest fit to the JWST NIRSpec phase curve observations. Therefore, we choose to favor the reflection-only eclipse depth based on the non-equilibrium cloud-free model ($cfne$) in our geometric albedo calculations. 

To derive a geometric albedo $A_g$, we express the eclipse depth as 
\begin{equation}
\delta_{ecl,r} = A_g \left(\frac{R_p}{a}\right)^2
\end{equation}
where $R_p$ and $a$ are planetary radius and semimajor axis (see Table \ref{table:params_table}). Using the favored $cfne$ model, we find a geometric albedo of \redt{$0.053 \pm 0.03$}. We note that the uncertainty reported on this value only considers the uncertainty in our own eclipse depth measurement; uncertainty introduced in modeling and removing emission from our phase curve is not considered, and we acknowledge may be significant, especially due to the limited number of GCMs considered. That said, our reported geometric albedo is consistent with the upper limit of $0.087$ reported in \citet{scandariato2022}, derived through a similar process. We also compute a geometric albedo from the $cfce$ and $cldne$ models, which yield an $A_g = 0.047\pm0.03$ and $A_g = 0.050\pm0.03$ respectively (see Table \ref{table:results_table}). 

\subsubsection{JWST NIRSpec emission}
\label{subsubsec:JWST NIRSpec emission}

By basing our removed TESS emission on observed IR emission from JWST NIRSpec measurements, we can compute a geometric albedo result with uncertainties from both TESS and JWST observations. However, this method requires the assumption that the temperature of the photosphere observed by TESS and JWST are the same. The temperatures of these photospheric layers depend on atmospheric composition, and they may not be the same across the wavelengths observed. Due to this possible discrepancy, it is likely that uncertainties on the results from this method are also underestimated. 

Using the JWST NIRSpec phase curve result from \citet{challener2024}, we find an eclipse depth of $3289 \pm 11~$ppm. This emission-dominated eclipse depth is consistent with a Spitzer IRAC $3.6\mu m$ depth of $3231\pm60~$ppm reported in \citet{stevenson2017}. Modeling the planet as a blackbody emitter, and using a PHOENIX \citep{husser2013} model spectrum for the host star, we compute an estimated emission-only eclipse depth of $25.7\pm0.3$ ppm in the TESS bandpass. Removing this modeled emission component from our observed TESS eclipse depth suggests a reflection-only eclipse depth of \redt{$104\pm34$ ppm}, and a geometric albedo \redt{$A_g = 0.09\pm0.03$}. This albedo, though higher than we find with our previous analysis, is consistent with the work of \citet{arora2024}, which obtains a value of $A_g = 0.109\pm0.087$ by fitting model spectra to HST and Spitzer observations of WASP-43b. That said, due to the uncertainties inherent in our analytical approach, we do not claim this to be a statistically significant geometric albedo measurement, and our result is generally consistent with the conclusion that this planet lacks reflective clouds and hazes at the atmospheric pressures probed by our measurements.

\subsection{Phase Curve Mapping}
\label{subsec: phase curve mapping}
Our mapping results indicate an eastward phase curve offset of \redt{$44\pm18$ degrees}. This result is consistent with the optical phase curve offset of $~50$ degrees eastward reported in \citet{scandariato2022}. As discussed, our TESS phase curve is sensitive to both reflected light and thermal emission. WASP-43b has also been studied using JWST MIRI and NIRSpec, which both probe thermal emission with negligible reflected-light contributions. Our measured TESS phase curve offset exceeds infrared offset measurements by \redt{$1.9\sigma$} or greater: MIRI mapping analysis shows an eastward peak flux offset of $7.5 \pm 0.5$ \citet{hammond2024}, while NIRSpec analysis in \citet{challener2024} finds an eastward offset of $10 \pm 0.8$ deg. 

\begin{figure}[ht!]
\centering
   \includegraphics[width=3.25in]{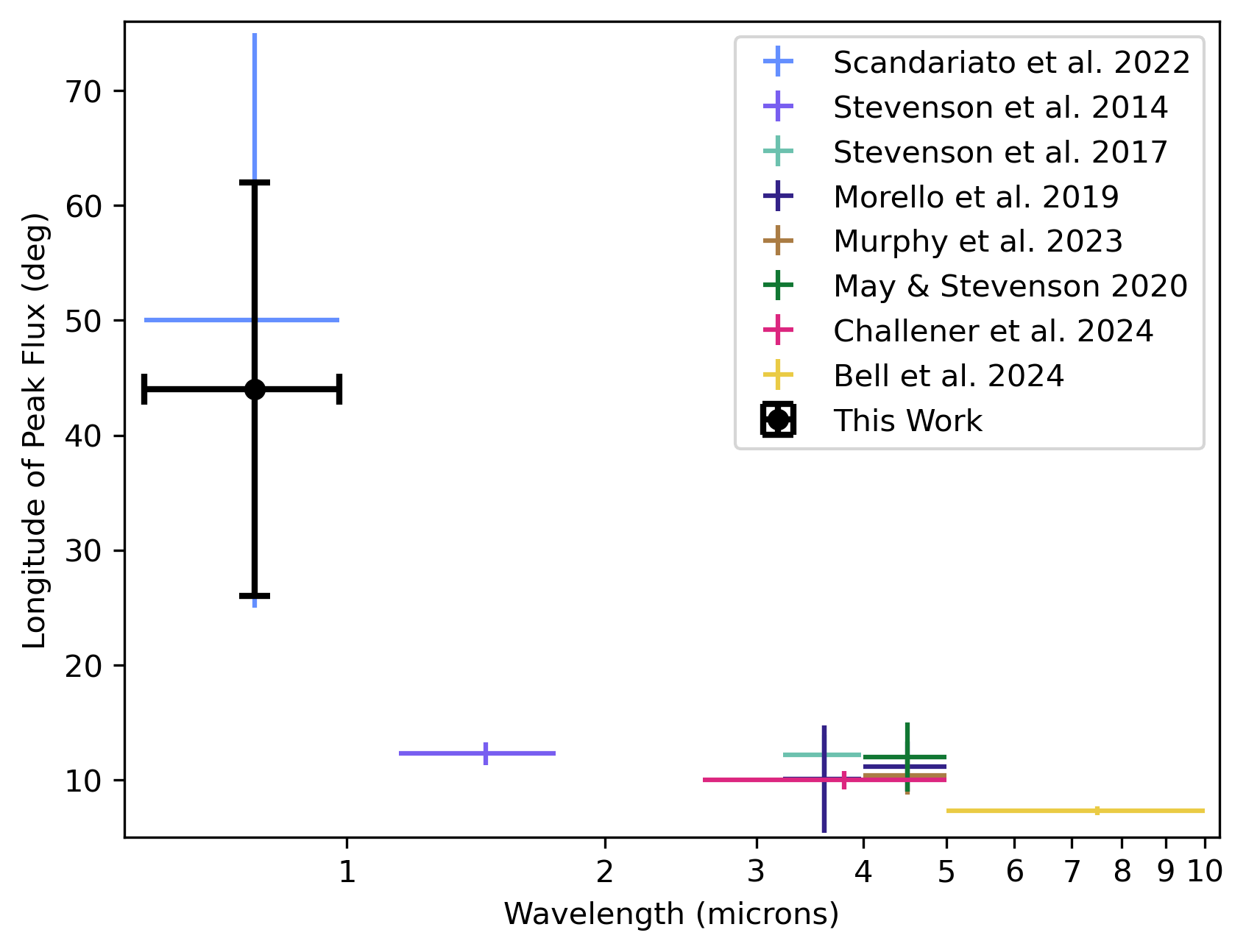}
   \caption{Wavelength dependence of WASP-43b peak flux offsets, with comparisons to previous analyses using data from TESS \citep{scandariato2022}, HST \citep{stevenson2014}, Spitzer \citep{stevenson2017, morello2019, murphy2023, may_stevenson2020}, MIRI \citep{bell2024}, and NIRSpec \citep{challener2024}. All positive offsets are east of the substellar point. The strong wavelength dependence of the peak flux offset, with lower wavelengths showing higher offsets, indicates sensitivity to differing dynamical processes across vertical layers of the atmosphere.}
   \label{fig:lit_vals}
\end{figure}

The strong wavelength dependence of the peak flux offset on WASP-43b’s dayside, as shown in Figure \ref{fig:lit_vals}, indicates sensitivity to differing dynamic processes across vertical layers of the atmosphere. \redt{Despite the difficulties in concretely disentangling emission and reflection at TESS wavelengths, our TESS phase offset measurement is not adequately explained by a dominant reflected light contribution.} If reflective clouds were present on WASP-43b's dayside, the peak phase offset found from TESS observations would be shifted westward \citep{parmentier2021} relative to hotspot offsets found in analyses of infrared observations. Instead we find the opposite: our relatively high peak phase offset result, paired with our relatively low dayside albedo, support a conclusion that a lack of dayside clouds allow TESS observations to probe deeper into the atmosphere \citep[e.g,][]{gressier2025}. \redt{At these deeper atmospheric pressure levels, longer radiative timescales allow zonal winds to transport heat further from the substellar point, resulting in a large eastward phase offset \citep[e.g.][]{showman2002, showman2008, showman2009}.}

A comparison of the three GCMs explored in Figure \ref{fig:lc} \redt{may support this conclusion as well: the phase curve offsets of all three models fall within $1\sigma$ of our measured TESS offset, with the cloudy model as the least similar, falling $0.5\sigma$ below our measured value.}  
In cloudless model scenarios, TESS wavelengths are sensitive to the deeper photosphere, which in turn are expected to have greater eastward phase curve offsets \citep[e.g.,][]{showman2009}. \redt{We note that although the GCMs considered in this work do support a cloud-free interpretation, this comparison alone does not exclude other possible cloud configurations, as we are referencing a limited set of model scenarios.}

\section{Conclusion} \label{sec:conclusion}
This work complements previous efforts \citep{stevenson2014, stevenson2017, morello2019, may_stevenson2020, scandariato2022, murphy2023, bell2024, hammond2024, challener2024} to map the dayside flux distribution of WASP-43b in a range of bandpasses. Pairing observations in emission- and reflection-sensitive bands allows a more informed approach to discussions of 3D dynamic processes and cloud/haze formation. In addition, the application of a robust eclipse mapping analysis to TESS data ensures physically plausible conclusions. 

We measure an eclipse depth of $130\pm34$ppm, a $3.8\sigma$ eclipse detection, which is in significant excess of our GCM emission model estimates. This measurement indicates that the observed phase curve includes excess flux beyond the planetary emission observed at IR wavelengths \citep{stevenson2014,stevenson2017, challener2024, hammond2024, bell2024}.

Using two independent approaches to model and remove the thermal emission contribution, we derive geometric albedo estimates in the range \redt{$A_g \approx 0.047-0.098$}, suggesting low reflectivity on the dayside. Although the removal of the thermal contribution is highly uncertain and model dependent, our range of values from both approaches is consistent with a lack of reflective clouds and hazes at the pressures probed by TESS observations. 

Our phase-curve mapping results indicate a phase curve offset \redt{$44\pm18$} degrees eastward of the substellar point, consistent with previous analysis in reflection-sensitive wavelength bands \citep{scandariato2022}, and exceeding the modest eastward peak flux offsets detected in redder emission-dominated bands \citep{stevenson2014,stevenson2017, challener2024, hammond2024, bell2024}. Reflective clouds and hazes can complicate the contribution functions relating wavelength to atmospheric pressure level and block optical observations from probing emission from the deep atmosphere \citep{dobbs_dixon2017}. A lack of reflective clouds and hazes at the pressures probed by TESS, as indicated by our results, may allow a substantial thermal emission component to dominate the phase curve offset. 

\redt{Our low measured albedo combined with the large eastward phase-curve offset favors the interpretation that the TESS band primarily probes a hot deep atmosphere in WASP-43b.} Existing literature \citep{scandariato2022, arora2024} and the albedo estimates in this work do not support the presence of a highly reflective cloud layer, indicating that the strong eastward peak flux offset in TESS bands is due to zonal heat circulation instead of longitudinally varying cloud coverage. The relative weakness of peak flux offsets in infrared measurements suggests more-effective heat recirculation (as would be expected in the deep atmosphere) at the TESS photosphere. A similar hot interior has been proposed for WASP-17b \citep{gressier2025} to explain excess shortwave planetary brightness. Furthermore, WASP-43b shows a lack of CH$_4$ on the nightside \citep{bell2024} that is also consistent with a hot interior \citep{yu2026}. Joint spectral analysis of observations across wavelength regimes could further clarify the atmospheric depths probed by each bandpass, serving to disentangle degeneracies \redt{and provide more definitive observational constraints on the possible hot interior of WASP-43b.}

This work highlights the utility of long-baseline survey data in hot-Jupiter mapping studies, and demonstrates the importance of combining optical and infrared observations to probe atmospheric structure in multiple dimensions. Extending this type of analysis to other TESS targets will provide a broader understanding of the wavelength-dependence of phase curve features. Comparing conclusions from multiple hot Jupiters can offer insight into population level trends in geometric albedo, vertical thermal structure, and possibly lead to evidence of partial dayside clouds and their longitudinal extent.

\begin{acknowledgments}
M. Lally acknowledges support by NASA under award No. 80NSSC25M7095.
R.C. Challener acknowledges support from NASA grant 80NSSC25K7152.
E.K.H. Lee acknowledges support from the CSH through the Bernoulli Fellowship.
\end{acknowledgments}




%
\facilities{TESS, JWST, Spitzer, HST}

\software{NumPy \citep{HarrisEtal2020natNumPy}, Matplotlib \citep{plt}, SciPy \citep{VirtanenEtal2020natmSciPy}, Scikit-learn \citep{PedregosaEtal2011jmlrScikitLearn}, starry \citep{starry}, ThERESA \citep{challener2022}, MC3 \citep{cubillos2017}, pysynphot \citep{pysynphot}, EXO-FMS \citep{lee2021}, gCMCRT \citep{lee2021_2}, lightkurve \citep{lightkurve-package}, astropy \citep{astropy2018}}




\bibliography{ref-2}{}
\bibliographystyle{aasjournalv7}



\end{document}